\documentclass[floatfix,twocolumn,showpacs,prl,amsmath]{revtex4}
\usepackage{graphicx}
\usepackage{bm}
\usepackage{amssymb}
\usepackage{dcolumn}
\usepackage{subfigure}
\usepackage{threeparttable}
\usepackage{multirow}
\usepackage{color}

\begin{document}
\newcommand{\vn}[1]{{\bf{#1}}}
\newcommand{\vht}[1]{{\boldsymbol{#1}}}
\newcommand{\matn}[1]{{\bf{#1}}}
\newcommand{\matnht}[1]{{\boldsymbol{#1}}}
\newcommand{\bege}{\begin{equation}}
\newcommand{\ee}{\end{equation}}
\newcommand{\bal}{\begin{aligned}}
\newcommand{\defbar}{\overline}
\newcommand{\SM}{\scriptstyle}
\newcommand{\eal}{\end{aligned}}
\newcommand{\udot}{\overset{.}{u}}
\newcommand{\exponential}[1]{{\exp(#1)}}
\newcommand{\phandot}[1]{\overset{\phantom{.}}{#1}}
\newcommand{\phandag}{\phantom{\dagger}}
\newcommand{\Trace}{\text{Tr}}
\title{Berry phase theory of Dzyaloshinskii-Moriya interaction and spin-orbit torques 
}
\author{Frank Freimuth}
\email[Corresp.~author:~]{f.freimuth@fz-juelich.de}
\author{Stefan Bl\"ugel}
\author{Yuriy Mokrousov}
\affiliation{Peter Gr\"unberg Institut and Institute for Advanced Simulation,
Forschungszentrum J\"ulich and JARA, 52425 J\"ulich, Germany}
\begin{abstract}
Recent experiments on current-induced domain wall motion in chiral magnets suggest
important contributions both from spin-orbit torques (SOTs) and from the Dzyaloshinskii-Moriya
interaction (DMI). We derive a Berry phase expression for the DMI
and show that within this Berry phase theory DMI and SOTs are intimately related, in a way  
formally analogous to the relation between orbital magnetization (OM) and anomalous Hall effect (AHE). We 
introduce the concept of the \textit{twist torque moment}, which probes the internal twist
of wave packets in chiral magnets in a similar way like the orbital moment probes the wave packet's
internal self rotation. We propose to interpret the Berry phase theory of DMI as a theory of
\textit{spiralization} in analogy to the modern theory of OM. We show that
the twist torque moment and the spiralization together give rise to a Berry phase governing the
response of the SOT to thermal gradients, in analogy to the intrinsic anomalous Nernst effect. 
The Berry phase theory of DMI is computationally very efficient because it only needs the electronic 
structure of the collinear magnetic system as input. As an application of the formalism 
we compute the DMI in Pt/Co, Pt/Co/O and Pt/Co/Al magnetic trilayers and show that the DMI is
highly anisotropic in these systems.  
\end{abstract}

\maketitle
Broken inversion symmetry in chiral magnets, such as B20 compounds, (Ga,Mn)As 
and asymmetric bi- or trilayers
opens new perspectives for current-induced
magnetization control via so-called spin-orbit torques 
(SOTs)~\cite{chernyshov_2009,manchon_zhang_2009,torque_macdonald,
CoPtAlO_spin_torque_rashba_Gambardella,brataas_stt_helimagnets}.
Notably, magnetization switching by SOTs in single collinear ferromagnetic layers
has been demonstrated experimentally~\cite{CoPtAlO_perpendicular_switching_Gambardella,
current_induced_switching_using_spin_torque_from_spin_hall_buhrman}.
In addition to SOTs also the Dzyaloshinskii-Moriya interaction (DMI) 
arises from the interplay of broken inversion
symmetry and spin-orbit interaction (SOI) in magnetic systems~\cite{dmi_moriya,dmi_dzyalo}.
Recent experiments and simulations suggest
that both SOTs and DMI substantially influence current-induced domain-wall 
motion in chiral magnets~\cite{simulations_dmi_walls_thiaville,
chiral_domain_wall_motion_parkin,chiral_domain_wall_motion_beach} and that their
combination may lead to a very efficient coupling of domain-wall motion to the applied current. 
Additionally, relations between SOTs and DMI have been proposed theoretically based on model
calculations~\cite{sot_dmi_stiles}.

Expanding the micromagnetic free energy density $F(\vn{r})$  at position $\vn{r}$ 
in terms of gradients $\partial \hat{\vn{n}}/\partial r_{j}$ of 
magnetization direction $\hat{\vn{n}}(\vn{r})$, we obtain in first order of the gradients
\bege\label{eq_first_order_free_energy}
F^{(1)}(\vn{r})=
\sum_{j}
\vn{D}_{j}(\hat{\vn{n}}(\vn{r}))
\cdot
\left(
\hat{\vn{n}}(\vn{r})\times\frac{\partial \hat{\vn{n}}(\vn{r})}{\partial r_{j}}
\right),
\ee
where
the Dzyaloshiskii vectors $\vn{D}_{j}(\hat{\vn{n}})$ will generally depend on 
magnetization direction $\hat{\vn{n}}(\vn{r})$.
Within \textit{ab initio} density-functional theory (DFT) methods, DMI is often 
computed by adding SOI
perturbatively to spirals with 
finite wave vectors $q$
and extracting $\vn{D}_{j}$ from the $q$-linear term in the 
dispersion E(q)~\cite{heide_dmi_few,heide_dmi_mnw,dmi_spirals_first_principles_heide}. 
Alternative methods for the calculation of DMI are based on multiple-scattering 
theory~\cite{dmi_weinberger,dmi_ebert} or a tight-binding representation of the 
electronic structure~\cite{dmi_katsnelson}. 

In the present work we develop a Berry phase theory of DMI. Our approach is based on 
expanding the free energy in terms of small gradients of magnetization direction within 
quantum mechanical perturbation theory. Formally, our Berry phase theory closely resembles the
quantum theory of OM. It drastically reduces the
computational burden, because it allows for calculating the DMI based on 
the collinear electronic structure. Additionally, 
the dependence of $\vn{D}_{j}(\hat{\vn{n}})$ on
magnetization direction $\hat{\vn{n}}$ is readily available, which is an advantage
for general magnetic structures whenever $\vn{D}_{j}(\hat{\vn{n}})$ is strongly anisotropic.  
Moreover, the relationship between SOT and DMI becomes visible 
within the Berry phase theory. It turns out that DMI and SOT are related in a similar way 
like OM and AHE are related. 
We introduce the concept of a twist torque moment of a given band, 
which turns out to be analogous to the orbital moment of a band in OM theory 
and propose to interpret DMI as a spiralization, i.e., a twist torque moment per volume, analogous to
the concept of magnetization as a magnetic moment per volume. We investigate how thermal gradients
and gradients in the chemical potential can give rise to SOTs and find that both the twist torque moment
and the DMI spiralization contribute to the SOT driven by statistical forces. 
Thus, DMI and SOT, the two effects
which can make the coupling between domain wall motion and applied current highly efficient,
are obtained from a common basis within the Berry phase theory.  
Finally, we apply our new method to the calculation of DMI in Pt/Co, Pt/Co/O and Pt/Co/Al magnetic thin
trilayer films and find that DMI is strongly anisotropic in these systems.

In the following we derive an expression for the calculation of $\vn{D}_{j}$ from the 
collinear electronic structure. We set $\hat{\vn{n}}=(\sin(\gamma),0,\cos(\gamma))$ and
consider small sinusoidal spatial oscillations of the angle $\gamma$ around 
zero, i.e., $\gamma(\vn{r})=\eta\sin(\vn{q}\cdot \vn{r})$, 
where $\eta$ is the smallness parameter.
Up to first order in $\eta$ we 
have $\hat{\vn{n}}(\vn{r})=(\eta\sin(\vn{q}\cdot \vn{r}),0,1)$.
Inserting this oscillating magnetization direction into Eq.~\eqref{eq_first_order_free_energy}, 
we obtain a spatially oscillating free energy density 
\bege
F^{(1)}(\vn{r})=\eta\sum_{j}
\vn{D}_{j}(\hat{\vn{e}}_{z})\cdot \hat{\vn{e}}_{y}q_{j}\cos(\vn{q}\cdot \vn{r}),
\ee
where $\hat{\vn{e}}_{y}$ and $\hat{\vn{e}}_{z}$ are unit vectors 
along the $y$ and $z$ directions, respectively. 
In order to extract $\vn{D}_{j}$ from this oscillating free energy density,
we multiply by $\cos(\vn{q}\cdot\vn{r})$
and integrate over the volume $V$:
\bege
\lim_{\vn{q}\to 0}
\frac{2}{V\eta}\frac{\partial}{\partial q_j}\int_{V}F^{(1)}(\vn{r}) \cos(\vn{q}\cdot\vn{r})d^3 r
=\vn{D}_{j}(\hat{\vn{e}}_{z})\cdot \hat{\vn{e}}_{y}.
\ee

The exchange interaction term in the LDA Hamiltonian is given 
by $\mu_{\rm B}B^{\rm xc}(\vn{r})\hat{\vn{n}}(\vn{r})\cdot\vht{\sigma}$, where
$\mu_{\rm B}$ is Bohr's magneton, $\vht{\sigma}$ is the vector of Pauli spin matrices, and
$B^{\rm xc}(\vn{r})$ is the exchange field. The oscillations of the magnetization direction
$\hat{\vn{n}}(\vn{r})$ perturb the wave functions of the collinear system.
To first order in the smallness parameter $\eta$ the perturbation operator is given by
$\delta V(\vn{r})=\mu_{\rm B}B^{\rm xc}(\vn{r})\sigma_{x}\eta\sin(\vn{q}\cdot\vn{r})$.
Below, we will first evaluate
\bege\label{eq_kamiltonian}
\begin{aligned}
K^{(1)}&=\frac{2}{\mathcal{N}}
\sum_{\vn{k}n}
f_{\vn{k}n}
\times\\
&\times\Re\int_{V}
(\psi_{\vn{k}n}(\vn{r}))^*(H_{0}-\mu N) \delta\psi_{\vn{k}n}(\vn{r}) \cos(\vn{q}\cdot\vn{r})d^3 r,
\end{aligned}
\ee
where $\psi_{\vn{k}n}(\vn{r})=e^{i\vn{k}\cdot\vn{r}}u_{\vn{k}n}(\vn{r})$ is the unperturbed Bloch function 
of band $n$ 
at $k$-point $\vn{k}$,
$\delta\psi_{\vn{k}n}(\vn{r})$ is the change of $\psi_{\vn{k}n}(\vn{r})$
within first order perturbation theory due to the perturbation $\delta V(\vn{r})$, 
$f_{\vn{k}n}=f(\mathcal{E}_{\vn{k}n})$ with Fermi function $f$ 
and $\mathcal{E}_{\vn{k}n}$ the unperturbed band energy, 
$H_{0}$ is the unperturbed LDA Hamiltonian of the collinear system, 
$\mu$ the chemical potential, 
$N$ the particle number operator, 
and $\mathcal{N}$ the number of $k$ points.  

Based on the relation $\frac{\partial}{\partial\beta}(\beta F)=E-\mu N$ between 
free energy and grand-canonical energy, where $\beta=(k_{\rm B}T)^{-1}$, we can relate $K^{(1)}$ and
$\vn{D}_{j}$ as follows:
\bege\label{eq_d_from_k}
\lim_{\vn{q}\to 0}
\frac{2}{V\eta}\frac{\partial}{\partial q_j}K^{(1)}
=
\frac{\partial}{\partial\beta}(\beta \vn{D}_{j}(\hat{\vn{e}}_{z})\cdot \hat{\vn{e}}_{y}).
\ee
 
The evaluation of Eq.~\eqref{eq_kamiltonian} is very similar to the derivation
of the quantum theory of OM~\cite{shi_quantum_theory_orbital_mag}.
Using the first-order perturbation theory expression for $\delta\psi_{\vn{k}n}(\vn{r})$ 
and switching from Bloch functions $\psi_{\vn{k}n}(\vn{r})$ to their lattice periodic parts
$u_{\vn{k}n}(\vn{r})$ we obtain 
\bege
\begin{aligned}
K^{(1)}&=\frac{\mu_{\rm B}\eta}{4\mathcal{N}}
\sum_{\vn{k}nm}
\left[
\mathcal{E}_{\vn{k}n}
+
\mathcal{E}_{\vn{k}+\vn{q}m}
-2\mu
\right]
\left[
f_{\vn{k}n}-f_{\vn{k}+\vn{q}m}
\right]
\times\\
&\times\Im\left[
\frac{
\langle
u_{\vn{k}n}|u_{\vn{k}+\vn{q}m}
\rangle
\langle
u_{\vn{k}+\vn{q}m}
|
B^{\rm xc}(\vn{r})\sigma_{x}
|
u_{\vn{k}n}
\rangle
}
{\mathcal{E}_{\vn{k}n} - \mathcal{E}_{\vn{k}+\vn{q}m} }
\right].
\end{aligned}
\ee

Differentiating with respect to $q_{j}$, taking the 
limit \mbox{$\vn{q}\rightarrow 0$} as prescribed by Eq.~\eqref{eq_d_from_k}, and generalizing to
arbitrary direction $\hat{\vn{n}}$ we arrive at the
expression
\bege\label{eq_dmi_berry}
\begin{aligned}
&\frac{\partial}{\partial\beta}(\beta \vn{D}_{j}(\hat{\vn{n}})\cdot \hat{\vn{e}}_{i})=\\
&\frac{\hbar}{\mathcal{N}V}
\sum_{\vn{k}}
\sum_{m\neq n}
\left[
\mathcal{E}_{\vn{k}n}-\mu
\right]
\left[
f_{\vn{k}n}
-
f_{\vn{k}m}
\right]
\times\\
&\times\Im
\left[ 
\frac{
\langle u_{\vn{k}n}  |\mathcal{T}_{i}| u_{\vn{k}m}  \rangle 
\langle u_{\vn{k}m}  |v_{j}(\vn{k})| u_{\vn{k}n}  \rangle
}
{(\mathcal{E}_{\vn{k}m}-\mathcal{E}_{\vn{k}n})^{2}}
\right]\\
&+
\frac{\hbar}{\mathcal{N}V}
\sum_{\vn{k}}
\sum_{m\neq n}
\left[
\mathcal{E}_{\vn{k}n}-\mu
\right]
f'(\mathcal{E}_{\vn{k}n})
\times\\
&\times\Im
\left[ 
\frac{
\langle u_{\vn{k}n}  |\mathcal{T}_{i}| u_{\vn{k}m}  \rangle 
\langle u_{\vn{k}m}  |v_{j}(\vn{k})| u_{\vn{k}n}  \rangle
}
{\mathcal{E}_{\vn{k}m}-\mathcal{E}_{\vn{k}n}}
\right],
\end{aligned}
\ee 
where $v_{j}(\vn{k})=
\frac{1}{\hbar}
\frac{\partial}{\partial k_j}
[e^{-i\vn{k}\cdot\vn{r}}H_{0}e^{i\vn{k}\cdot\vn{r}}]$
is the $j$-component of the velocity operator in crystal momentum representation, and
the torque operator at position
$\vn{r}$ is given by
$\vht{\mathcal{T}}(\vn{r})=\vn{m}\times\vn{B}^{\rm xc}(\vn{r})$ in terms of the 
spin magnetic moment operator $\vn{m}=-\mu_{\rm B}\vht{\sigma}$ and the exchange field $\vn{B}^{\rm xc}(\vn{r})$ and
$\mathcal{T}_{i}$ is its $i$ component. 

Integrating Eq.~\eqref{eq_dmi_berry} and 
defining $D_{ij}(\hat{\vn{n}})=\vn{D}_{j}(\hat{\vn{n}})\cdot \hat{\vn{e}}_{i}$
yields
\bege\label{eq_dmi_finite_temperature}
D_{ij}
=
\frac{1}{\mathcal{N}V}
\sum_{\vn{k}n}\big\{f_{\vn{k}n}A_{\vn{k}nij}
+\frac{1}{\beta}\ln [1+e^{-\beta(\mathcal{E}_{\vn{k}n}-\mu)}]
B_{\vn{k}nij}
\big\},
\ee
where 
\bege\label{eq_akn_kubo}
A_{\vn{k}nij}=\hbar\sum_{m\neq n}\Im
\left[
\frac{
\langle u_{\vn{k}n}  |\mathcal{T}_{i}| u_{\vn{k}m}  \rangle 
\langle u_{\vn{k}m}  |v_{j}(\vn{k})| u_{\vn{k}n}  \rangle
}
{
\mathcal{E}_{\vn{k}m}-\mathcal{E}_{\vn{k}n}
}
\right]
\ee
and
\bege\label{eq_bkn_kubo}
\begin{aligned}
B_{\vn{k}nij}
=-2\hbar\sum_{m\neq n}\Im
\left[
\frac{
\langle u_{\vn{k}n}  |\mathcal{T}_{i}| u_{\vn{k}m}  \rangle 
\langle u_{\vn{k}m}  |v_{j}(\vn{k})| u_{\vn{k}n}  \rangle
}
{
(\mathcal{E}_{\vn{k}m}-\mathcal{E}_{\vn{k}n})^2
}
\right].
\end{aligned}
\ee

Using
\bege\label{eq_torque_operator}
\vht{\mathcal{T}}=\frac{\partial H_0}{\partial \theta}\hat{\vn{e}}_{\phi}
-
\frac{1}{\sin\theta}\frac{\partial H_0}{\partial \phi}\hat{\vn{e}}_{\theta},
\ee
where $\theta$ and $\phi$ specify $\hat{\vn{n}}$ in spherical coordinates, i.e., 
$\hat{\vn{n}}=(\sin\theta\cos\phi,\sin\theta\sin\phi,\cos\theta)$, and
$\hat{\vn{e}}_{\theta}=\partial \hat{\vn{n}}/\partial\theta$,
$\hat{\vn{e}}_{\phi}=(1/\sin\theta)\partial \hat{\vn{n}}/\partial\phi$, we obtain
alternative expressions of $A_{\vn{k}n}$ and $B_{\vn{k}n}$ in terms of 
derivatives of the wave functions with
respect to both crystal momentum $\vn{k}$ and the spherical coordinates of the 
magnetization direction:
\bege\label{eq_akn_angular_derivatives}
\begin{aligned}
A_{\vn{k}nij}=-
\hat{\vn{e}}_{i} \cdot
\left[
\hat{\vn{e}}_{\phi}
\Im
\left\langle
\frac{\partial u_{\vn{k}n}}{\partial\theta}
\right|
(\mathcal{E}_{\vn{k}n}-H_{0}(\vn{k}))
\left|
\frac{\partial u_{\vn{k}n}}{\partial k_{j}}
\right\rangle \right. &\\
\left. -\hat{\vn{e}}_{\theta} \frac{1}{\sin\theta}
\Im
\left\langle
\frac{\partial u_{\vn{k}n}}{\partial\phi}
\right|
(\mathcal{E}_{\vn{k}n}-H_{0}(\vn{k}))
\left|
\frac{\partial u_{\vn{k}n}}{\partial k_{j}}
\right\rangle
\right]&,
\end{aligned}
\ee
and
\bege
\begin{aligned}
B_{\vn{k}nij}\label{eq_bkn_angular_derivatives}
=
-2
\hat{\vn{e}}_{i} \cdot
\left[
\hat{\vn{e}}_{\phi}
\Im
\left\langle
\frac{\partial u_{\vn{k}n}}{\partial\theta}
\left|
\frac{\partial u_{\vn{k}n}}{\partial k_{j}}\right.
\right\rangle \right.&\\
\left. -\hat{\vn{e}}_{\theta} \frac{1}{\sin\theta}
\Im
\left\langle
\frac{\partial u_{\vn{k}n}}{\partial\phi}
\left|
\frac{\partial u_{\vn{k}n}}{\partial k_{j}}\right.
\right\rangle
\right]&,
\end{aligned}
\ee
where $H_{0}(\vn{k})=e^{-i\vn{k}\cdot\vn{r}}H_{0}e^{i\vn{k}\cdot \vn{r}}$ is the crystal momentum
representation of the Hamiltonian $H_{0}$.
We note that the special case of $\theta=0$, in which case $\sin\theta=0$ in the
numerators, is obtained from these
equations by considering a finite $\theta$ and then taking the limit of $\theta\rightarrow 0$.
The case of $\theta=\pi$ is treated similarly. Alternatively, one may avoid the problems
at $\theta=0$ and $\theta=\pi$ by choosing the polar axis such that $\theta$ is between 
0 and $\pi$. 
At zero temperature Eq.~\eqref{eq_dmi_finite_temperature} becomes
\bege\label{eq_dmi_zero_temp}
\begin{aligned}
D_{ij}=\frac{1}{\mathcal{N}V}
\sum_{\vn{k}n}
f_{\vn{k}n}
[
A_{\vn{k}nij}-(\mathcal{E}_{\vn{k}n}-\mu)B_{\vn{k}nij}
]\quad\quad\quad\quad\quad\quad&\\
=-
\frac{\hat{\vn{e}}_{i}}{\mathcal{N}V} \cdot
\left[
\hat{\vn{e}}_{\phi}
\Im
\left\langle
\frac{\partial u_{\vn{k}n}}{\partial\theta}
\right|
(2\mu-\mathcal{E}_{\vn{k}n}-H_{0}(\vn{k}))
\left|
\frac{\partial u_{\vn{k}n}}{\partial k_{j}}
\right\rangle \right.\,\,\,& \\
\left. -\frac{\hat{\vn{e}}_{\theta}}{\sin\theta}
\Im
\left\langle
\frac{\partial u_{\vn{k}n}}{\partial\phi}
\right|
(2\mu-\mathcal{E}_{\vn{k}n}-H_{0}(\vn{k}))
\left|
\frac{\partial u_{\vn{k}n}}{\partial k_{j}}
\right\rangle
\right]&.
\end{aligned}
\ee
Comparing Eq.~\eqref{eq_dmi_finite_temperature} with the quantum theory of 
OM~\cite{shi_quantum_theory_orbital_mag,berry_anomalous_thermoelectric_xiao}
one finds strong formal analogies, where $B_{\vn{k}nij}$ corresponds to the Berry curvature
$i\langle\nabla_{\vn{k}}u_{\vn{k}n}|\times|\nabla_{\vn{k}}u_{\vn{k}n}\rangle$ and $A_{\vn{k}nij}$ 
corresponds to the orbital 
moment $(e/2\hbar)i\langle\nabla_{\vn{k}}u_{\vn{k}n}|\times[\mathcal{E}_{\vn{k}n}-H_0]|\nabla_{\vn{k}}u_{\vn{k}n}\rangle$ of state $n$. 
Therefore, we define the \textit{twist torque moments} of state $n$ by $A_{\vn{k}nij}$. 
In perfect analogy to the theory of OM
the DMI at zero temperature given by Eq.~\eqref{eq_dmi_zero_temp}  is not simply the sum of the 
twist torque moments of all occupied states divided by the volume
but there is a Berry curvature correction due to $B_{\vn{k}nij}$.
In order to counterpart the formal analogies on the level of terminology, it is tempting to call $D_{ij}$ the
\textit{DMI spiralization}.  

Considering \textit{finite} systems rather than infinite periodic crystals we can further substantiate
the analogies between OM and DMI spiralization.  
For finite systems, 
the orbital magnetic moment $\vn{p}_{\rm orb}$
can be expressed via the moment $\vn{r}\times \vn{j}(\vn{r})$
of the current density $\vn{j}(\vn{r})$ as 
\bege
\vn{p}_{\rm orb}=\frac{1}{2}
\sum_{n}f_{n}
\int d^3 r
(\psi_{n}(\vn{r}))^{*}
\vn{r}\times \vn{j}(\vn{r})
\psi_{n}(\vn{r})
. 
\ee 
Since the position operator $\vn{r}$ is not compatible with periodic boundary 
conditions, it is replaced either by the velocity operator or by derivatives with 
respect to crystal momentum in the theory of OM for 
infinite systems whenever the formalism is based on the 
Bloch-periodic eigenfunctions of the Hamiltonian~\cite{resta_review_om,orbital_magnetization_periodic_insulators,
orbital_magnetization_crystalline_solids,theory_orbital_magnetization_disordered_systems}.
In this situation the OM is expressed in terms of Berry phases.
Likewise, it is natural to interpret DMI in finite systems in terms of the 
moments $r_{i}\mathcal{T}_{j}(\vn{r})$ of the torque, i.e.,
\bege
D_{ij}=\frac{1}{V}
\sum_{n}f_{n}
\int d^3 r
(\psi_{n}(\vn{r}))^{*}
r_{j}\mathcal{T}_{i}(\vn{r})
\psi_{n}(\vn{r})
,
\ee 
because if the magnetization rotates by an angle $\delta\Phi$ around the axis $\vn{s}$, the
associated energy change is $\vht{\mathcal{T}}\cdot\vn{s}\delta\Phi$. Thus,
in this picture of DMI the free energy change due to magnetization gradients arises from the
asymmetry of the torque, which can be quantified by its moments. However, like in the
case of the orbital moment, an expression of DMI involving the position operator
cannot be used for infinite systems with periodic boundary conditions and the correct
theory has instead of the position operator either velocity 
operators (see Eq.~\eqref{eq_akn_kubo} and Eq.~\eqref{eq_bkn_kubo}) 
or derivatives with respect to crystal momentum
(see Eq.~\eqref{eq_akn_angular_derivatives} and 
Eq.~\eqref{eq_bkn_angular_derivatives}).

While Thonhauser et al.~\cite{orbital_magnetization_periodic_insulators} 
and Shi et al.~\cite{shi_quantum_theory_orbital_mag} derived the expressions
of the OM quantum mechanically rigorously for Bloch electrons in 
crystalline solids, exactly the same expressions have been obtained from
semiclassical wave-packet dynamics~\cite{berry_anomalous_thermoelectric_xiao}. 
At first glance astonishing, this agreement results from the semiclassical theory getting
exact as the length scale of the perturbation goes to infinity~\cite{shi_quantum_theory_orbital_mag}. 
This limit $\vn{q}\rightarrow 0$ 
is also taken explicitly in our definition of the DMI spiralization in Eq.~\eqref{eq_d_from_k}.
Thus, the expressions obtained for DMI within the semiclassical formalism have to 
reproduce our results. It will become clear in the following that developing the semiclassical picture of
SOT and DMI is quite rewarding.

To get started semiclassically, we define the twist torque moment of a wave 
packet $|W_{\vn{k}n}\rangle$ constructed from band $n$ with 
average crystal momentum $\vn{k}$ by
\bege
A_{\vn{k}nij}^{W}=\langle W_{\vn{k}n}|(r_{j}-r^{W}_{\vn{k}nj})\mathcal{T}_{i}|W_{\vn{k}n}\rangle,
\ee
where $r^{W}_{\vn{k}nj}=\langle W_{\vn{k}n}|r_{j}|W_{\vn{k}n} \rangle$ are the coordinates of the center of $|W_{\vn{k}n}\rangle$. 
In Ref.~\cite{wave_packets_sundaram} the 
details of the construction of $|W_{\vn{k}n}\rangle$ from Bloch functions 
are given and a wave packet formalism is presented there which allows for rewriting
wave packet expectation values in terms of Berry phase expressions.  
In the case of the twist torque moment we obtain
\bege
A^{W}_{\vn{k}nij}=A_{\vn{k}nij}.
\ee
Thus, our previous expression for the twist torque moment 
of band $n$ in Eq.~\eqref{eq_akn_angular_derivatives} is equivalent to the expression of
the twist torque moment of the wave packet constructed from band $n$ if the crystal momentum
$\vn{k}$ in Eq.~\eqref{eq_akn_angular_derivatives} is identified with the mean wave 
vector of the wave packet. In the theory of SOT and DMI the twist torque moment plays the same
role as the orbital moment does in the theory of AHE and OM. It is associated with the internal twist of
the wave packet which locally prefers a noncollinear spiral magnetic structure such that any enforcement 
of magnetic collinearity leads to torques countering this collinearity, the moment of which is the twist torque moment.

Inclusion of the SOT into the picture marks the next stage of our semiclassical expedition.
If an external electric field $\vn{E}$ is applied to the system, a torque $\vn{T}$ arises, which
is given within linear response by $\vn{T}=\vn{t}\vn{E}$, where $\vn{t}$ is the torkance tensor. As we
have shown recently~\cite{ibcsoit}, 
the intrinsic even torkance $t_{ij}^{\rm even}(\hat{\vn{n}})=(t_{ij}(\hat{\vn{n}})
+t_{ij}(-\hat{\vn{n}}))/2$ can be expressed in terms of the Berry 
curvature Eq.~\eqref{eq_bkn_angular_derivatives} as
\bege 
t_{ij}^{\rm even}(\hat{\vn{n}})=-\frac{e}{\mathcal{N}}\sum_{\vn{k}n}f_{\vn{k}n}B_{\vn{k}nij},
\ee
where $e>0$ is the elementary positive charge.
Thus, the electron in band $n$ with crystal momentum $\vn{k}$ exerts the torque
\bege\label{eq_torque_expression}
\vn{T}_{\vn{k}n}=\frac{\partial\mathcal{E}_{\vn{k}n}}{\partial \theta}\hat{\vn{e}}_{\phi}
-
\frac{1}{\sin\theta}\frac{\partial\mathcal{E}_{\vn{k}n}}{\partial \phi}\hat{\vn{e}}_{\theta}
-\sum_{ij}eB_{\vn{k}nij}\hat{\vn{e}}_{i}E_{j}
\ee 
on the magnetization. 
The first two terms are simply the expectation value of
the torque operator Eq.~\eqref{eq_torque_operator}
and give rise to the magnetocrystalline
anisotropy energy. The last term is the even SOT. 
This equation should be compared to the wave packet's semiclassical equation of motion 
\bege\label{eq_rdot_expression}
\frac{d \vn{r}^{W}_{\vn{k}n}}{d t}=\frac{1}{\hbar}\frac{\partial \mathcal{E}_{\vn{k}n}}{\partial \vn{k}}
+\frac{e}{\hbar}\vn{E}\times\vht{\Omega}_{\vn{k}n},
\ee
where the first term is the group velocity and the last term is the anomalous velocity due to the
Berry curvature $\vht{\Omega}_{\vn{k}n}$ which gives rise to the AHE.
We find that the even SOT, related to the Berry curvature $B_{\vn{k}nij}$, is analogous to the 
AHE, related to the Berry curvature $\vht{\Omega}_{\vn{k}n}$. In metallic systems
the application of an external electric field leads to additional responses of the system besides
the ones due to these Berry curvature terms, because the
Fermi sphere is shifted and the states are occupied according to a nonequilibrium distribution 
function. Evaluated for such a nonequilibrium distribution, the first two terms in Eq.~\eqref{eq_torque_expression} yield the odd torque $T_{ij}^{\rm odd}(\hat{\vn{n}})=(T_{ij}(\hat{\vn{n}})
-T_{ij}(-\hat{\vn{n}}))/2$~\cite{ibcsoit}, while the first term in Eq.~\eqref{eq_rdot_expression}  
gives rise to the normal electrical transport current.  

According to the Einstein relation, a gradient in the chemical potential $\nabla\mu$ has 
the same effects as
an applied electric field $\vn{E}=\nabla\mu/e$. However, in the absence of an applied electric field
the rightmost terms in Eq. ~\eqref{eq_torque_expression} and in Eq.~\eqref{eq_rdot_expression}
vanish. Therefore, the question arises how the Berry phases enter the theory when statistical 
rather than mechanical forces drive the electrons. 
Xiao et al.~\cite{berry_anomalous_thermoelectric_xiao} have shown that
in order to solve this puzzle in the case of Eq.~\eqref{eq_rdot_expression}, it is 
important to distinguish between local currents and transport currents and to take the
wave packet's finite spread into account in the equation for the local current.
The derivation of a \textit{local torque} in analogy to the local current of Xiao et al.\ is rather
straightforward. Due to the twist torque moment gradients of the temperature or of the chemical
potential lead to a correction term for the local torque, which is given by
\bege
\delta T^{\rm loc}_{i}=
\frac{1}{\mathcal{N}}\sum_{\vn{k}n}
\sum_{j}\frac{\partial}{\partial r_j}
f_{\vn{k}n}(\vn{r})A_{\vn{k}nij},
\ee 
where $f_{\vn{k}n}(\vn{r})$ depends on $\vn{r}$ due to gradients in temperature or in chemical potential. 
Thus, a Berry phase term, namely $A_{\vn{k}nij}$, manifests itself in the local torque 
whenever temperature or chemical potential are inhomogeneous across the sample, because the 
twist torque moment couples to these gradients, whereby it is partly converted into a torque. 
Next, we have to subtract the gradients of the DMI spiralization to obtain the measurable torque. This
step is analogous to the subtraction of the curl of magnetization from the local current density in
the work of Xiao et al.\ and leads to the following correction term to the measurable torque 
in the presence of gradients of T or $\mu$: 
\bege\label{eq_statistical_force_correction_term}
\delta T_{i}=
-\frac{1}{\mathcal{N}}\sum_{\vn{k}nj}B_{\vn{k}nij}
\frac{\partial}{\partial r_j}
\frac{1}{\beta}
\ln[1+e^{-\beta(\mathcal{E}_{\vn{k}n}-\mu)}].
\ee

From Eq.~\eqref{eq_statistical_force_correction_term} one easily obtains the torque due to a
chemical potential gradient $\nabla\mu$:
\bege
\delta T_{i}=-\frac{e}{\mathcal{N}}\sum_{\vn{k}nj}
f_{\vn{k}n}B_{\vn{k}nij}\frac{1}{e}\frac{\partial \mu}{\partial r_j}
=\sum_{j}t_{ij}^{\rm even}\frac{1}{e}\frac{\partial \mu}{\partial r_j},
\ee
showing that the Einstein relation is satisfied. 
The torque due to a temperature gradient can be written as
\bege
\delta T_{i}=\sum_{j}\frac{1}{e}\frac{\partial T}{\partial r_{j}}
\int d\mathcal{E}\frac{d f_{\vn{k}n}}{d\mu}t^{\rm even}_{ij}(\mathcal{E})|_{T=0}\frac{\mathcal{E}-\mu}{T},
\ee
where $t^{\rm even}_{ij}(\mathcal{E})|_{T=0}$ is the even torkance at zero temperature with 
Fermi energy set to $\mathcal{E}$. Thus, analogously to the 
intrinsic anomalous Nernst effect~\cite{berry_anomalous_thermoelectric_xiao}, the proper definition
of local and measurable torques introduces the Berry phases into the response to thermal gradients. 

We proceed to formulate DMI in terms of Green functions.
Thereby, our objective is twofold. First, the Green function formulation will
allow us to connect DMI and SOT from a different perspective. 
Second, we expect that a Green function theory 
for DMI will sometimes be favorable, e.g.\ for the investigation of DMI in disordered systems. 
Using the residue theorem we can prove the identity
\bege
\begin{aligned}
&\Im
\int
d\mathcal{E} f(\mathcal{E})
\frac{1}{\mathcal{E}-\mathcal{E}_q+i0^+}
\frac{1}{(\mathcal{E}-\mathcal{E}_p+i0^+)^2}
=\\
&=\pi
\left[
\frac{[f(\mathcal{E}_p)-f(\mathcal{E}_q)]}{(\mathcal{E}_p-\mathcal{E}_q)^2}
+
\frac{f'(\mathcal{E}_p)}{\mathcal{E}_q-\mathcal{E}_p}
\right],
\end{aligned}
\ee
which allows us to rewrite Eq.~\eqref{eq_dmi_berry} in terms of Green functions as follows:
\begin{gather}\label{eq_dmi_green_functions_hamil}
\begin{aligned}
&\frac{\partial}{\partial \beta}(\beta D_{ij})=\frac{-1}{2\pi V\hbar^2}\Im\int
d\mathcal{E} f(\mathcal{E})\times \\
\times{\rm Tr}\langle
&\mathcal{T}_{i}G^{\rm R}(\mathcal{E})v_{j}
G^{\rm R}(\mathcal{E})(H_0-\mu)G^{\rm R}(\mathcal{E})\\
-
&\mathcal{T}_{i}
G^{\rm R}(\mathcal{E})(H_0-\mu)G^{\rm R}(\mathcal{E})
v_{j}G^{\rm R}(\mathcal{E})
\rangle,
\end{aligned}
\end{gather}
where $G^{\rm R}(\mathcal{E})=[\mathcal{E}-H_0+i0^+]^{-1}$ is the
retarded Green function. Making use of 
\begin{gather}\label{eq_identity}
\begin{aligned}
0&=\Im\int
d\mathcal{E} f(\mathcal{E})\times \\
\times{\rm Tr}\langle
&\mathcal{T}_{i}G^{\rm R}(\mathcal{E})v_{j}
G^{\rm R}(\mathcal{E})(H_0-\mathcal{E})G^{\rm R}(\mathcal{E})\\
-
&\mathcal{T}_{i}
G^{\rm R}(\mathcal{E})(H_0-\mathcal{E})G^{\rm R}(\mathcal{E})
v_{j}G^{\rm R}(\mathcal{E})
\rangle
\end{aligned}
\end{gather}
we can simplify Eq.~\eqref{eq_dmi_green_functions_hamil} into the form
\begin{gather}\label{eq_dmi_green_functions}
\begin{aligned}
\frac{\partial}{\partial \beta}&(\beta D_{ij})=\frac{1}{hV}\Re\int
d\mathcal{E} f(\mathcal{E})(\mathcal{E}-\mu)\times\\
&\times{\rm Tr}\langle
\mathcal{T}_{i}G^{\rm R}(\mathcal{E})v_{j}
\frac{dG^{\rm R}(\mathcal{E})}{d\mathcal{E}}
-
\mathcal{T}_{i}\frac{dG^{\rm R}(\mathcal{E})}{d\mathcal{E}}v_{j}G^{\rm R}(\mathcal{E})
\rangle.
\end{aligned}\raisetag{2\baselineskip}
\end{gather}

This result can be directly compared with the SOT torkance $t_{ij}$, which is a sum of three contributions~\cite{ibcsoit},
i.e., $t_{ij}=t^{\rm I(a)}_{ij}+t^{\rm I(b)}_{ij}+t^{\rm II}_{ij}$, where
\begin{gather}\label{eq_kubo_linear_response}
\begin{aligned}
t^{\rm I(a)\phantom{I}}_{ij}\!\!\!\!=-\frac{e}{h}\int&
d\mathcal{E}
\frac{d f(\mathcal{E})}{d \mathcal{E}}
\phantom{\Re}
{\rm Tr}
\langle\mathcal{T}_{i}
G^{\rm R}(\mathcal{E})
v_{j}
G^{\rm A}(\mathcal{E})
\rangle
\\
t^{\rm I(b)\phantom{I}}_{ij}\!\!\!\!=\phantom{-}\frac{e}{h}\int
&d\mathcal{E}\frac{d f(\mathcal{E})}{d \mathcal{E}}
{\Re}
{\rm Tr}
\langle\mathcal{T}_{i}
G^{\rm R}(\mathcal{E})
v_{j}
G^{\rm R}(\mathcal{E})
\rangle\phantom{(1)}
\\
t^{\rm II\phantom{(a)}}_{ij}\!\!\!\!=\phantom{-}\frac{e}{h}\int
&d\mathcal{E} f(\mathcal{E})
\quad\!\!
{\Re}{\rm Tr}\langle
\mathcal{T}_{i}G^{\rm R}(\mathcal{E})v_{j}
\frac{dG^{\rm R}(\mathcal{E})}{d\mathcal{E}}\\
 &\quad\quad\quad\quad\quad\,-
\mathcal{T}_{i}\frac{dG^{\rm R}(\mathcal{E})}{d\mathcal{E}}v_{j}G^{\rm R}(\mathcal{E})
\rangle,
\end{aligned}\raisetag{4\baselineskip}
\end{gather}
with
$G^{\rm A}(\mathcal{E})$ the  
advanced Green function.
Clearly, $\frac{\partial}{\partial \beta}(\beta D_{ij})$ differs from $t^{\rm II}_{ij}$
only by an additional factor $(\mathcal{E}-\mu)/e$ in the integrand 
as well as a factor $1/V$ which can
be left away whenever energy per unit cell is the desired unit of energy density.
It is interesting that only part of the torkance, namely only $t^{\rm II}_{ij}$, is related to
the $D_{ij}$. But in fact also for the torkance itself, $t^{\rm II}_{ij}$ plays a special role, because in 
contrast to the other two terms it is a Fermi sea integral. Moreover, its complex 
version -- i.e., without taking the real part -- can be analytically continued into the upper half complex plane.
Additionally, $t^{\rm II}_{ij}$ contributes only to the even SOT torkance but not to the odd one.
Since DMI is a ground state property then so is $t^{\rm II}_{ij}$. The complete SOT torkance is generally not a
ground state property but a transport one, in particular the intraband contribution to the odd torkance involves 
the relaxation times~\cite{ibcsoit}.

Defining the energy-resolved torkance $\vartheta_{ij}(E)$ at $T=0$ as
\begin{gather}\label{eq_energy_resolved_torkance}
\begin{aligned}
\vartheta_{ij}(E)=&\frac{e}{h}
\delta(\mathcal{E}-\mu)
\phantom{\Re}
{\rm Tr}
\langle\mathcal{T}_{i}
G^{\rm R}(\mathcal{E})
v_{j}
G^{\rm A}(\mathcal{E})
\rangle
\\
-&\frac{e}{h}
\delta(\mathcal{E}-\mu)
{\Re}
{\rm Tr}
\langle\mathcal{T}_{i}
G^{\rm R}(\mathcal{E})
v_{j}
G^{\rm R}(\mathcal{E})
\rangle\phantom{(1)}
\\
+&\frac{e}{h} \theta(\mu-\mathcal{E})
{\Re}{\rm Tr}\langle
\mathcal{T}_{i}G^{\rm R}(\mathcal{E})v_{j}
\frac{dG^{\rm R}(\mathcal{E})}{d\mathcal{E}}\\
 &\quad\quad\quad\quad\quad\,-
\mathcal{T}_{i}\frac{dG^{\rm R}(\mathcal{E})}{d\mathcal{E}}v_{j}G^{\rm R}(\mathcal{E})
\rangle,
\end{aligned}
\end{gather}
we may write at $T=0$:
\bege\label{eq_sot_dmi_0k}
\begin{aligned}
t_{ij}=&\int d\mathcal{E}\vartheta_{ij}(\mathcal{E})\\
D_{ij}=&\frac{1}{eV}\int d\mathcal{E}(\mathcal{E}-\mu)\vartheta_{ij}(\mathcal{E}).
\end{aligned}
\ee
Since $\delta(\mathcal{E}-\mu)(\mathcal{E}-\mu)=0$, the first two lines in Eq.~\eqref{eq_energy_resolved_torkance} do not contribute to $D_{ij}$.
Eq.~\eqref{eq_sot_dmi_0k} suggests that $t_{ij}$ and $D_{ij}$ will generally behave similarly, in particular
from the symmetry point of view. Doing a symmetry analysis for $\vartheta_{ij}$ is sufficient to determine
the symmetries of both $t_{ij}$ and $D_{ij}$. In systems with strongly anisotropic 
SOT~\cite{symmetry_spin_orbit_torques}
we expect also the DMI to be anisotropic as a consequence of Eq.~\eqref{eq_sot_dmi_0k}. 
Furthermore, sign and magnitude of $t_{ij}$ and $D_{ij}$ will often be correlated.

We turn now to the computational aspects of the Berry phase formalism of DMI spiralisation.
Recently, a Wannier function based method for the calculation 
of the orbital moment directly from the Berry phase 
expressions has been presented~\cite{wannier_based_calculation_orbital_magnetization_crystals}.
Conceptually, the direct evaluation of observables from their Berry phase
representation (e.g.\  Eq.~\eqref{eq_akn_angular_derivatives} and 
Eq.~\eqref{eq_bkn_angular_derivatives}) is very appealing and appears to be advantageous
over the use of the corresponding Kubo formula 
expressions  (e.g.\  Eq.~\eqref{eq_akn_kubo} and 
Eq.~\eqref{eq_bkn_kubo}). However, in the case of DMI, the curvatures involve also
angular derivatives. They are similar to phase-space Berry phases, or mixed real-space
momentum-space Berry phases~\cite{rsms}. This opens an interesting practical aspect 
for the generalization of
the Wannier function concept to higher dimensions by performing Fourier transforms not only with respect to
crystal momentum, but also with respect to other parameters of the Hamiltonian, such as the
magnetic structure parameters. In order to compute the DMI in disordered systems, Eq.~\eqref{eq_dmi_green_functions} should generally be suitable. 
Since Eq.~\eqref{eq_dmi_green_functions} involves only the retarded Green function, the energy integration
can be performed in the upper half complex plane, analogously to the calculation of the charge density from the
Green function in Green function based DFT codes.   
Irrespective of whether the explicit Berry phase based or the wave function based or the Green function based version of the
DMI theory is used, one will in most cases achieve computational speed-ups compared to extracting DMI from 
spin spiral calculations due to several reasons: The symmetry of 
the collinear system is higher than the symmetry of the noncollinear spin spiral system. 
If spin-orbit interaction is treated within second variation, the computational time demand is dominated by the
diagonalization of the collinear Hamiltonians, which can be faster by a factor of 4 compared to diagonalizing the noncollinear Hamiltonian. In systems with anisotropic DMI sampling the DMI vector with the Berry phase method is more efficient
than reconstructing the information from spiral calculations which by construction average over the various 
magnetization directions comprising the spiral. 

As an application of the Berry phase method we compute DMI in Pt/Co, Pt/Co/O and Pt/Co/Al thin films
composed of 10 atomic layers
of Pt(111), 3 atomic layers of hcp Co and one additional atomic layer of O or Al. 
Our Pt/Co, Pt/Co/O and Pt/Co/Al thin films are realistic models of trilayer structures such as
Pt/Co/AlO$_{x}$ and Pt/Co/MgO currently studied extensively experimentally due to SOT and due to the combination
of SOT and DMI to allow for highly efficient current induced domain wall motion.
The computational details of the DFT electronic structure calculations are given in Ref.~\cite{ibcsoit}, in which 
the authors studied the SOT in these systems. 
We constructed maximally localized
Wannier functions from the relativistic first-principles Bloch functions  
in order to evaluate Eq.~\eqref{eq_dmi_zero_temp}, Eq.~\eqref{eq_akn_kubo} 
and Eq.~\eqref{eq_bkn_kubo} computationally efficiently by making
use of the Wannier interpolation technique~\cite{WannierPaper,wannier90,rmp_wannier90}.
For $\hat{\vn{n}}$ perpendicular to the films (i.e., along $z$ direction), 
the computed DMI spiralizations $D_{yx}=-D_{xy}$ 
of Pt/Co, Pt/Co/O and Pt/Co/Al are respectively
11.3, 15.0 and 20.7 meV\r{A} per unit cell (one unit cell contains 3 Co atoms). The in-plane unit cell area is
6.65 \r{A}$^2$. 
Interestingly, it has been found that the OM converges much faster with respect to the density of the
$k$-mesh than the AHE, even though both are computed from similar 
Berry phase expressions~\cite{wannier_based_calculation_orbital_magnetization_crystals}. We report an 
analogous observation
for the DMI spiralization: Using uniform 32x32, 64x64, 128x128 and 512x512 $k$-meshes we obtain 
spiralizations $D_{yx}$ of 12.8, 11.8, 12.2, and 11.3 meV\r{A} per unit cell for Pt/Co. This suggests the option of
doing quick estimates of $D_{ij}$ using coarse $k$ grids. Such estimates could also be done without Wannier interpolation
directly within the first principles codes.     
Computing $D_{yx}$ for various directions of $\hat{\vn{n}}$, 
we find the DMI to depend strongly on the direction of $\hat{\vn{n}}$. 
E.g.\ for $\hat{\vn{n}}$ along $x$ direction, $D_{yx}$ is smaller by a factor of 3 than for 
$\hat{\vn{n}}$ in perpendicular to film direction in the case of the Pt/Co/Al film. Anisotropies of this order of magnitude are
typical of transport coefficients in non-cubic crystals. Inclusion of such anisotropy terms of the DMI into
the micromagnetic energy functionals used for simulation of current-induced domain-wall motion in chiral
magnets is therefore expected to affect results on the quantitative level.   

In conclusion we showed that DMI can be formulated in terms of a Berry phase theory.
We derived this Berry phase theory by expanding the free energy functional within rigorous
quantum mechanical perturbation theory in terms of gradients of magnetization direction. 
Formally, our Berry phase theory closely resembles the
quantum theory of OM and drastically reduces the
computational burden, because it allows for calculating the DMI based on 
the collinear electronic structure. 
We worked out the analogies between OM and DMI and showed that
the orbital moment of a band is counterparted by a twist torque moment within the Berry phase DMI theory.
This twist torque moment is of the same fundamental importance for the intrinsic even SOT driven by
thermal gradients like the orbital moment of a band is for the intrinsic anomalous Nernst effect.
We investigated the formal relations between DMI and SOT and found them to be the same as those 
between OM and AHE. We propose to interpret DMI as a spiralization, i.e., a twist torque moment
per volume, in analogy to the magnetization, which can be interpreted as a magnetic moment per volume. 
Besides formulating the DMI explicitly as a Berry phase theory we also derived various equivalent alternative 
expressions which can be conveniently implemented within first principles DFT codes, including expressions in terms of
Green functions, which allow the computation of DMI in disordered systems. 
As a practical application of the formalism we computed the DMI in Pt/Co, Pt/Co/O and Pt/Co/Al thin films. We found
the DMI to be strongly anisotropic in these systems.

We gratefully acknowledge discussions with 
R.~Bamler,
G.~Bihlmayer,  
P.~Gambardella, K.~Garello, G.~Gaudin, 
M.~Heide, E.~Ju\'e, I.~M.~Miron, 
A.~Rosch
and A.~Thiaville,
computing time on the supercomputers \mbox{JUQUEEN} 
and \mbox{JUROPA} 
at J\"ulich Supercomputing Center and funding under the HGF-YIG programme VH-NG-513.


\begin{thebibliography}{33}
\expandafter\ifx\csname natexlab\endcsname\relax\def\natexlab#1{#1}\fi
\expandafter\ifx\csname bibnamefont\endcsname\relax
  \def\bibnamefont#1{#1}\fi
\expandafter\ifx\csname bibfnamefont\endcsname\relax
  \def\bibfnamefont#1{#1}\fi
\expandafter\ifx\csname citenamefont\endcsname\relax
  \def\citenamefont#1{#1}\fi
\expandafter\ifx\csname url\endcsname\relax
  \def\url#1{\texttt{#1}}\fi
\expandafter\ifx\csname urlprefix\endcsname\relax\def\urlprefix{URL }\fi
\providecommand{\bibinfo}[2]{#2}
\providecommand{\eprint}[2][]{\url{#2}}

\bibitem[{\citenamefont{Chernyshov et~al.}(2009)\citenamefont{Chernyshov,
  Overby, Liu, Furdyna, Lyanda-Geller, and Rokhinson}}]{chernyshov_2009}
\bibinfo{author}{\bibfnamefont{A.}~\bibnamefont{Chernyshov}},
  \bibinfo{author}{\bibfnamefont{M.}~\bibnamefont{Overby}},
  \bibinfo{author}{\bibfnamefont{X.}~\bibnamefont{Liu}},
  \bibinfo{author}{\bibfnamefont{J.~K.} \bibnamefont{Furdyna}},
  \bibinfo{author}{\bibfnamefont{Y.}~\bibnamefont{Lyanda-Geller}},
  \bibnamefont{and} \bibinfo{author}{\bibfnamefont{L.~P.}
  \bibnamefont{Rokhinson}}, \bibinfo{journal}{Nature Phys.}
  \textbf{\bibinfo{volume}{5}}, \bibinfo{pages}{656} (\bibinfo{year}{2009}).

\bibitem[{\citenamefont{Manchon and Zhang}(2009)}]{manchon_zhang_2009}
\bibinfo{author}{\bibfnamefont{A.}~\bibnamefont{Manchon}} \bibnamefont{and}
  \bibinfo{author}{\bibfnamefont{S.}~\bibnamefont{Zhang}},
  \bibinfo{journal}{Phys. Rev. B} \textbf{\bibinfo{volume}{79}},
  \bibinfo{pages}{094422} (\bibinfo{year}{2009}).

\bibitem[{\citenamefont{Garate and MacDonald}(2009)}]{torque_macdonald}
\bibinfo{author}{\bibfnamefont{I.}~\bibnamefont{Garate}} \bibnamefont{and}
  \bibinfo{author}{\bibfnamefont{A.~H.} \bibnamefont{MacDonald}},
  \bibinfo{journal}{Phys. Rev. B} \textbf{\bibinfo{volume}{80}},
  \bibinfo{pages}{134403} (\bibinfo{year}{2009}).

\bibitem[{\citenamefont{Mihai~Miron et~al.}(2010)\citenamefont{Mihai~Miron,
  Gaudin, Auffret, Rodmacq, Schuhl, Pizzini, Vogel, and
  Gambardella}}]{CoPtAlO_spin_torque_rashba_Gambardella}
\bibinfo{author}{\bibfnamefont{I.}~\bibnamefont{Mihai~Miron}},
  \bibinfo{author}{\bibfnamefont{G.}~\bibnamefont{Gaudin}},
  \bibinfo{author}{\bibfnamefont{S.}~\bibnamefont{Auffret}},
  \bibinfo{author}{\bibfnamefont{B.}~\bibnamefont{Rodmacq}},
  \bibinfo{author}{\bibfnamefont{A.}~\bibnamefont{Schuhl}},
  \bibinfo{author}{\bibfnamefont{S.}~\bibnamefont{Pizzini}},
  \bibinfo{author}{\bibfnamefont{J.}~\bibnamefont{Vogel}}, \bibnamefont{and}
  \bibinfo{author}{\bibfnamefont{P.}~\bibnamefont{Gambardella}},
  \bibinfo{journal}{Nature Mater.} \textbf{\bibinfo{volume}{9}},
  \bibinfo{pages}{230} (\bibinfo{year}{2010}).

\bibitem[{\citenamefont{Hals and Brataas}(2013)}]{brataas_stt_helimagnets}
\bibinfo{author}{\bibfnamefont{K.~M.~D.} \bibnamefont{Hals}} \bibnamefont{and}
  \bibinfo{author}{\bibfnamefont{A.}~\bibnamefont{Brataas}},
  \bibinfo{journal}{Phys. Rev. B} \textbf{\bibinfo{volume}{87}},
  \bibinfo{pages}{174409} (\bibinfo{year}{2013}).

\bibitem[{\citenamefont{Mihai~Miron et~al.}(2011)\citenamefont{Mihai~Miron,
  Garello, Gaudin, Zermatten, Costache, Auffret, Bandiera, Rodmacq, Schuhl, and
  Gambardella}}]{CoPtAlO_perpendicular_switching_Gambardella}
\bibinfo{author}{\bibfnamefont{I.}~\bibnamefont{Mihai~Miron}},
  \bibinfo{author}{\bibfnamefont{K.}~\bibnamefont{Garello}},
  \bibinfo{author}{\bibfnamefont{G.}~\bibnamefont{Gaudin}},
  \bibinfo{author}{\bibfnamefont{P.-J.} \bibnamefont{Zermatten}},
  \bibinfo{author}{\bibfnamefont{M.~V.} \bibnamefont{Costache}},
  \bibinfo{author}{\bibfnamefont{S.}~\bibnamefont{Auffret}},
  \bibinfo{author}{\bibfnamefont{S.}~\bibnamefont{Bandiera}},
  \bibinfo{author}{\bibfnamefont{B.}~\bibnamefont{Rodmacq}},
  \bibinfo{author}{\bibfnamefont{A.}~\bibnamefont{Schuhl}}, \bibnamefont{and}
  \bibinfo{author}{\bibfnamefont{P.}~\bibnamefont{Gambardella}},
  \bibinfo{journal}{Nature} \textbf{\bibinfo{volume}{476}},
  \bibinfo{pages}{189} (\bibinfo{year}{2011}).

\bibitem[{\citenamefont{Liu et~al.}(2012)\citenamefont{Liu, Lee, Gudmundsen,
  Ralph, and
  Buhrman}}]{current_induced_switching_using_spin_torque_from_spin_hall_buhrma%
n}
\bibinfo{author}{\bibfnamefont{L.}~\bibnamefont{Liu}},
  \bibinfo{author}{\bibfnamefont{O.~J.} \bibnamefont{Lee}},
  \bibinfo{author}{\bibfnamefont{T.~J.} \bibnamefont{Gudmundsen}},
  \bibinfo{author}{\bibfnamefont{D.~C.} \bibnamefont{Ralph}}, \bibnamefont{and}
  \bibinfo{author}{\bibfnamefont{R.~A.} \bibnamefont{Buhrman}},
  \bibinfo{journal}{Phys. Rev. Lett.} \textbf{\bibinfo{volume}{109}},
  \bibinfo{pages}{096602} (\bibinfo{year}{2012}).

\bibitem[{\citenamefont{Moriya}(1960)}]{dmi_moriya}
\bibinfo{author}{\bibfnamefont{T.}~\bibnamefont{Moriya}},
  \bibinfo{journal}{Phys. Rev.} \textbf{\bibinfo{volume}{120}},
  \bibinfo{pages}{91} (\bibinfo{year}{1960}).

\bibitem[{\citenamefont{Dzyaloshinsky}(1958)}]{dmi_dzyalo}
\bibinfo{author}{\bibfnamefont{I.}~\bibnamefont{Dzyaloshinsky}},
  \bibinfo{journal}{Journal of Physics and Chemistry of Solids}
  \textbf{\bibinfo{volume}{4}}, \bibinfo{pages}{241} (\bibinfo{year}{1958}).

\bibitem[{\citenamefont{Thiaville et~al.}(2013)\citenamefont{Thiaville, Rohart,
  Jue, Cros, and Fert}}]{simulations_dmi_walls_thiaville}
\bibinfo{author}{\bibfnamefont{A.}~\bibnamefont{Thiaville}},
  \bibinfo{author}{\bibfnamefont{S.}~\bibnamefont{Rohart}},
  \bibinfo{author}{\bibfnamefont{E.}~\bibnamefont{Jue}},
  \bibinfo{author}{\bibfnamefont{V.}~\bibnamefont{Cros}}, \bibnamefont{and}
  \bibinfo{author}{\bibfnamefont{A.}~\bibnamefont{Fert}},
  \bibinfo{journal}{Nature Nanotech.} \textbf{\bibinfo{volume}{8}},
  \bibinfo{pages}{527} (\bibinfo{year}{2013}).

\bibitem[{\citenamefont{Thomas et~al.}(2013)\citenamefont{Thomas, Ryu, Yang,
  and Parkin}}]{chiral_domain_wall_motion_parkin}
\bibinfo{author}{\bibfnamefont{L.}~\bibnamefont{Thomas}},
  \bibinfo{author}{\bibfnamefont{K.}~\bibnamefont{Ryu}},
  \bibinfo{author}{\bibfnamefont{S.}~\bibnamefont{Yang}}, \bibnamefont{and}
  \bibinfo{author}{\bibfnamefont{S.~S.~P.} \bibnamefont{Parkin}},
  \bibinfo{journal}{Nature Nanotech.} \textbf{\bibinfo{volume}{8}},
  \bibinfo{pages}{527} (\bibinfo{year}{2013}).

\bibitem[{\citenamefont{Emori et~al.}(2013)\citenamefont{Emori, Bauer, Ahn,
  Martinez, and Beach}}]{chiral_domain_wall_motion_beach}
\bibinfo{author}{\bibfnamefont{S.}~\bibnamefont{Emori}},
  \bibinfo{author}{\bibfnamefont{U.}~\bibnamefont{Bauer}},
  \bibinfo{author}{\bibfnamefont{S.}~\bibnamefont{Ahn}},
  \bibinfo{author}{\bibfnamefont{E.}~\bibnamefont{Martinez}}, \bibnamefont{and}
  \bibinfo{author}{\bibfnamefont{G.~S.~D.} \bibnamefont{Beach}},
  \bibinfo{journal}{Nature Mater.} \textbf{\bibinfo{volume}{12}},
  \bibinfo{pages}{611} (\bibinfo{year}{2013}).

\bibitem[{\citenamefont{Kim et~al.}(2013)\citenamefont{Kim, Lee, Lee, and
  Stiles}}]{sot_dmi_stiles}
\bibinfo{author}{\bibfnamefont{K.-W.} \bibnamefont{Kim}},
  \bibinfo{author}{\bibfnamefont{H.-W.} \bibnamefont{Lee}},
  \bibinfo{author}{\bibfnamefont{K.-J.} \bibnamefont{Lee}}, \bibnamefont{and}
  \bibinfo{author}{\bibfnamefont{M.~D.} \bibnamefont{Stiles}}
  (\bibinfo{year}{2013}), \bibinfo{note}{arXiv:1308.1198}.

\bibitem[{\citenamefont{Heide et~al.}(2008)\citenamefont{Heide, Bihlmayer, and
  Bl\"ugel}}]{heide_dmi_few}
\bibinfo{author}{\bibfnamefont{M.}~\bibnamefont{Heide}},
  \bibinfo{author}{\bibfnamefont{G.}~\bibnamefont{Bihlmayer}},
  \bibnamefont{and} \bibinfo{author}{\bibfnamefont{S.}~\bibnamefont{Bl\"ugel}},
  \bibinfo{journal}{Phys. Rev. B} \textbf{\bibinfo{volume}{78}},
  \bibinfo{pages}{140403} (\bibinfo{year}{2008}).

\bibitem[{\citenamefont{Ferriani et~al.}(2008)\citenamefont{Ferriani, von
  Bergmann, Vedmedenko, Heinze, Bode, Heide, Bihlmayer, Bl\"ugel, and
  Wiesendanger}}]{heide_dmi_mnw}
\bibinfo{author}{\bibfnamefont{P.}~\bibnamefont{Ferriani}},
  \bibinfo{author}{\bibfnamefont{K.}~\bibnamefont{von Bergmann}},
  \bibinfo{author}{\bibfnamefont{E.~Y.} \bibnamefont{Vedmedenko}},
  \bibinfo{author}{\bibfnamefont{S.}~\bibnamefont{Heinze}},
  \bibinfo{author}{\bibfnamefont{M.}~\bibnamefont{Bode}},
  \bibinfo{author}{\bibfnamefont{M.}~\bibnamefont{Heide}},
  \bibinfo{author}{\bibfnamefont{G.}~\bibnamefont{Bihlmayer}},
  \bibinfo{author}{\bibfnamefont{S.}~\bibnamefont{Bl\"ugel}}, \bibnamefont{and}
  \bibinfo{author}{\bibfnamefont{R.}~\bibnamefont{Wiesendanger}},
  \bibinfo{journal}{Phys. Rev. Lett.} \textbf{\bibinfo{volume}{101}},
  \bibinfo{pages}{027201} (\bibinfo{year}{2008}).

\bibitem[{\citenamefont{Heide et~al.}(2009)\citenamefont{Heide, Bihlmayer, and
  Bl\"ugel}}]{dmi_spirals_first_principles_heide}
\bibinfo{author}{\bibfnamefont{M.}~\bibnamefont{Heide}},
  \bibinfo{author}{\bibfnamefont{G.}~\bibnamefont{Bihlmayer}},
  \bibnamefont{and} \bibinfo{author}{\bibfnamefont{S.}~\bibnamefont{Bl\"ugel}},
  \bibinfo{journal}{Physica B: Condensed Matter}
  \textbf{\bibinfo{volume}{404}}, \bibinfo{pages}{2678} (\bibinfo{year}{2009}).

\bibitem[{\citenamefont{Udvardi et~al.}(2003)\citenamefont{Udvardi, Szunyogh,
  Palot\'as, and Weinberger}}]{dmi_weinberger}
\bibinfo{author}{\bibfnamefont{L.}~\bibnamefont{Udvardi}},
  \bibinfo{author}{\bibfnamefont{L.}~\bibnamefont{Szunyogh}},
  \bibinfo{author}{\bibfnamefont{K.}~\bibnamefont{Palot\'as}},
  \bibnamefont{and}
  \bibinfo{author}{\bibfnamefont{P.}~\bibnamefont{Weinberger}},
  \bibinfo{journal}{Phys. Rev. B} \textbf{\bibinfo{volume}{68}},
  \bibinfo{pages}{104436} (\bibinfo{year}{2003}).

\bibitem[{\citenamefont{Ebert and Mankovsky}(2009)}]{dmi_ebert}
\bibinfo{author}{\bibfnamefont{H.}~\bibnamefont{Ebert}} \bibnamefont{and}
  \bibinfo{author}{\bibfnamefont{S.}~\bibnamefont{Mankovsky}},
  \bibinfo{journal}{Phys. Rev. B} \textbf{\bibinfo{volume}{79}},
  \bibinfo{pages}{045209} (\bibinfo{year}{2009}).

\bibitem[{\citenamefont{Katsnelson et~al.}(2010)\citenamefont{Katsnelson,
  Kvashnin, Mazurenko, and Lichtenstein}}]{dmi_katsnelson}
\bibinfo{author}{\bibfnamefont{M.~I.} \bibnamefont{Katsnelson}},
  \bibinfo{author}{\bibfnamefont{Y.~O.} \bibnamefont{Kvashnin}},
  \bibinfo{author}{\bibfnamefont{V.~V.} \bibnamefont{Mazurenko}},
  \bibnamefont{and} \bibinfo{author}{\bibfnamefont{A.~I.}
  \bibnamefont{Lichtenstein}}, \bibinfo{journal}{Phys. Rev. B}
  \textbf{\bibinfo{volume}{82}}, \bibinfo{pages}{100403}
  (\bibinfo{year}{2010}).

\bibitem[{\citenamefont{Shi et~al.}(2007)\citenamefont{Shi, Vignale, Xiao, and
  Niu}}]{shi_quantum_theory_orbital_mag}
\bibinfo{author}{\bibfnamefont{J.}~\bibnamefont{Shi}},
  \bibinfo{author}{\bibfnamefont{G.}~\bibnamefont{Vignale}},
  \bibinfo{author}{\bibfnamefont{D.}~\bibnamefont{Xiao}}, \bibnamefont{and}
  \bibinfo{author}{\bibfnamefont{Q.}~\bibnamefont{Niu}},
  \bibinfo{journal}{Phys. Rev. Lett.} \textbf{\bibinfo{volume}{99}},
  \bibinfo{pages}{197202} (\bibinfo{year}{2007}).

\bibitem[{\citenamefont{Xiao et~al.}(2006)\citenamefont{Xiao, Yao, Fang, and
  Niu}}]{berry_anomalous_thermoelectric_xiao}
\bibinfo{author}{\bibfnamefont{D.}~\bibnamefont{Xiao}},
  \bibinfo{author}{\bibfnamefont{Y.}~\bibnamefont{Yao}},
  \bibinfo{author}{\bibfnamefont{Z.}~\bibnamefont{Fang}}, \bibnamefont{and}
  \bibinfo{author}{\bibfnamefont{Q.}~\bibnamefont{Niu}},
  \bibinfo{journal}{Phys. Rev. Lett.} \textbf{\bibinfo{volume}{97}},
  \bibinfo{pages}{026603} (\bibinfo{year}{2006}).

\bibitem[{\citenamefont{Resta}(2010)}]{resta_review_om}
\bibinfo{author}{\bibfnamefont{R.}~\bibnamefont{Resta}}, \bibinfo{journal}{J.
  Phys.: Condens. Matter} \textbf{\bibinfo{volume}{22}},
  \bibinfo{pages}{123201} (\bibinfo{year}{2010}).

\bibitem[{\citenamefont{Thonhauser et~al.}(2005)\citenamefont{Thonhauser,
  Ceresoli, Vanderbilt, and Resta}}]{orbital_magnetization_periodic_insulators}
\bibinfo{author}{\bibfnamefont{T.}~\bibnamefont{Thonhauser}},
  \bibinfo{author}{\bibfnamefont{D.}~\bibnamefont{Ceresoli}},
  \bibinfo{author}{\bibfnamefont{D.}~\bibnamefont{Vanderbilt}},
  \bibnamefont{and} \bibinfo{author}{\bibfnamefont{R.}~\bibnamefont{Resta}},
  \bibinfo{journal}{Phys. Rev. Lett.} \textbf{\bibinfo{volume}{95}},
  \bibinfo{pages}{137205} (\bibinfo{year}{2005}).

\bibitem[{\citenamefont{Ceresoli et~al.}(2006)\citenamefont{Ceresoli,
  Thonhauser, Vanderbilt, and
  Resta}}]{orbital_magnetization_crystalline_solids}
\bibinfo{author}{\bibfnamefont{D.}~\bibnamefont{Ceresoli}},
  \bibinfo{author}{\bibfnamefont{T.}~\bibnamefont{Thonhauser}},
  \bibinfo{author}{\bibfnamefont{D.}~\bibnamefont{Vanderbilt}},
  \bibnamefont{and} \bibinfo{author}{\bibfnamefont{R.}~\bibnamefont{Resta}},
  \bibinfo{journal}{Phys. Rev. B} \textbf{\bibinfo{volume}{74}},
  \bibinfo{pages}{024408} (\bibinfo{year}{2006}).

\bibitem[{\citenamefont{Zhu et~al.}(2012)\citenamefont{Zhu, Yang, Fang, Liu,
  and Yao}}]{theory_orbital_magnetization_disordered_systems}
\bibinfo{author}{\bibfnamefont{G.}~\bibnamefont{Zhu}},
  \bibinfo{author}{\bibfnamefont{S.~A.} \bibnamefont{Yang}},
  \bibinfo{author}{\bibfnamefont{C.}~\bibnamefont{Fang}},
  \bibinfo{author}{\bibfnamefont{W.~M.} \bibnamefont{Liu}}, \bibnamefont{and}
  \bibinfo{author}{\bibfnamefont{Y.}~\bibnamefont{Yao}},
  \bibinfo{journal}{Phys. Rev. B} \textbf{\bibinfo{volume}{86}},
  \bibinfo{pages}{214415} (\bibinfo{year}{2012}).

\bibitem[{\citenamefont{Sundaram and Niu}(1999)}]{wave_packets_sundaram}
\bibinfo{author}{\bibfnamefont{G.}~\bibnamefont{Sundaram}} \bibnamefont{and}
  \bibinfo{author}{\bibfnamefont{Q.}~\bibnamefont{Niu}},
  \bibinfo{journal}{Phys. Rev. B} \textbf{\bibinfo{volume}{59}},
  \bibinfo{pages}{14915} (\bibinfo{year}{1999}).

\bibitem[{\citenamefont{Freimuth et~al.}(2013)\citenamefont{Freimuth, Bl\"ugel,
  and Mokrousov}}]{ibcsoit}
\bibinfo{author}{\bibfnamefont{F.}~\bibnamefont{Freimuth}},
  \bibinfo{author}{\bibfnamefont{S.}~\bibnamefont{Bl\"ugel}}, \bibnamefont{and}
  \bibinfo{author}{\bibfnamefont{Y.}~\bibnamefont{Mokrousov}}
  (\bibinfo{year}{2013}), \bibinfo{note}{arXiv:1305.4873}.

\bibitem[{\citenamefont{Garello et~al.}(2013)\citenamefont{Garello, Miron,
  Avci, Freimuth, Mokrousov, Bl\"ugel, Auffret, Boulle, Gaudin, and
  Gambardella}}]{symmetry_spin_orbit_torques}
\bibinfo{author}{\bibfnamefont{K.}~\bibnamefont{Garello}},
  \bibinfo{author}{\bibfnamefont{I.~M.} \bibnamefont{Miron}},
  \bibinfo{author}{\bibfnamefont{C.~O.} \bibnamefont{Avci}},
  \bibinfo{author}{\bibfnamefont{F.}~\bibnamefont{Freimuth}},
  \bibinfo{author}{\bibfnamefont{Y.}~\bibnamefont{Mokrousov}},
  \bibinfo{author}{\bibfnamefont{S.}~\bibnamefont{Bl\"ugel}},
  \bibinfo{author}{\bibfnamefont{S.}~\bibnamefont{Auffret}},
  \bibinfo{author}{\bibfnamefont{O.}~\bibnamefont{Boulle}},
  \bibinfo{author}{\bibfnamefont{G.}~\bibnamefont{Gaudin}}, \bibnamefont{and}
  \bibinfo{author}{\bibfnamefont{P.}~\bibnamefont{Gambardella}},
  \bibinfo{journal}{Nature Nanotech.} \textbf{\bibinfo{volume}{8}},
  \bibinfo{pages}{587} (\bibinfo{year}{2013}).

\bibitem[{\citenamefont{Lopez et~al.}(2012)\citenamefont{Lopez, Vanderbilt,
  Thonhauser, and
  Souza}}]{wannier_based_calculation_orbital_magnetization_crystals}
\bibinfo{author}{\bibfnamefont{M.~G.} \bibnamefont{Lopez}},
  \bibinfo{author}{\bibfnamefont{D.}~\bibnamefont{Vanderbilt}},
  \bibinfo{author}{\bibfnamefont{T.}~\bibnamefont{Thonhauser}},
  \bibnamefont{and} \bibinfo{author}{\bibfnamefont{I.}~\bibnamefont{Souza}},
  \bibinfo{journal}{Phys. Rev. B} \textbf{\bibinfo{volume}{85}},
  \bibinfo{pages}{014435} (\bibinfo{year}{2012}).

\bibitem[{\citenamefont{Bamler et~al.}(2013)\citenamefont{Bamler, Freimuth,
  Mokrousov, and Rosch}}]{rsms}
\bibinfo{author}{\bibfnamefont{R.}~\bibnamefont{Bamler}},
  \bibinfo{author}{\bibfnamefont{F.}~\bibnamefont{Freimuth}},
  \bibinfo{author}{\bibfnamefont{Y.}~\bibnamefont{Mokrousov}},
  \bibnamefont{and} \bibinfo{author}{\bibfnamefont{A.}~\bibnamefont{Rosch}}
  (\bibinfo{year}{2013}), \bibinfo{note}{arXiv:1307.8085}.

\bibitem[{\citenamefont{Freimuth et~al.}(2008)\citenamefont{Freimuth,
  Mokrousov, Wortmann, Heinze, and Bl\"ugel}}]{WannierPaper}
\bibinfo{author}{\bibfnamefont{F.}~\bibnamefont{Freimuth}},
  \bibinfo{author}{\bibfnamefont{Y.}~\bibnamefont{Mokrousov}},
  \bibinfo{author}{\bibfnamefont{D.}~\bibnamefont{Wortmann}},
  \bibinfo{author}{\bibfnamefont{S.}~\bibnamefont{Heinze}}, \bibnamefont{and}
  \bibinfo{author}{\bibfnamefont{S.}~\bibnamefont{Bl\"ugel}},
  \bibinfo{journal}{Phys. Rev. B} \textbf{\bibinfo{volume}{78}},
  \bibinfo{pages}{035120} (\bibinfo{year}{2008}).

\bibitem[{\citenamefont{Mostofi et~al.}(2008)\citenamefont{Mostofi, Yates, Lee,
  Souza, Vanderbilt, and Marzari}}]{wannier90}
\bibinfo{author}{\bibfnamefont{A.~A.} \bibnamefont{Mostofi}},
  \bibinfo{author}{\bibfnamefont{J.~R.} \bibnamefont{Yates}},
  \bibinfo{author}{\bibfnamefont{Y.-S.} \bibnamefont{Lee}},
  \bibinfo{author}{\bibfnamefont{I.}~\bibnamefont{Souza}},
  \bibinfo{author}{\bibfnamefont{D.}~\bibnamefont{Vanderbilt}},
  \bibnamefont{and} \bibinfo{author}{\bibfnamefont{N.}~\bibnamefont{Marzari}},
  \bibinfo{journal}{Computer Physics Communications}
  \textbf{\bibinfo{volume}{178}}, \bibinfo{pages}{685 } (\bibinfo{year}{2008}).

\bibitem[{\citenamefont{Marzari et~al.}(2012)\citenamefont{Marzari, Mostofi,
  Yates, Souza, and Vanderbilt}}]{rmp_wannier90}
\bibinfo{author}{\bibfnamefont{N.}~\bibnamefont{Marzari}},
  \bibinfo{author}{\bibfnamefont{A.~A.} \bibnamefont{Mostofi}},
  \bibinfo{author}{\bibfnamefont{J.~R.} \bibnamefont{Yates}},
  \bibinfo{author}{\bibfnamefont{I.}~\bibnamefont{Souza}}, \bibnamefont{and}
  \bibinfo{author}{\bibfnamefont{D.}~\bibnamefont{Vanderbilt}},
  \bibinfo{journal}{Rev. Mod. Phys.} \textbf{\bibinfo{volume}{84}},
  \bibinfo{pages}{1419} (\bibinfo{year}{2012}).

\end{thebibliography}

\end{document}